\documentclass[11pt,a4paper]{article}
\usepackage{cite}

\usepackage{amsmath, amsthm,mathtools,url,amssymb,upgreek,slashed,dsfont}

\usepackage{mathabx} 

\usepackage{xcolor}

\usepackage{float}

\usepackage{ifpdf}
\ifpdf
  \usepackage[pdftex]{graphicx}
  \usepackage{epstopdf}
\else
  \usepackage[dvips]{graphicx}
\fi
\parskip=\baselineskip

\numberwithin{equation}{section}

\newcommand{\beq}{\begin{equation}}
\newcommand{\eeq}{\end{equation}}

\newcommand{\RR}{\mathbb{R}}

\newcommand{\SL}{\mathrm{SL}}

\renewcommand{\sl}{\mathfrak{sl}}
\newcommand{\so}{\mathfrak{so}}

\renewcommand{\gg}{\mathfrak{g}}
\newcommand{\hh}{\mathfrak{h}}
\newcommand{\kk}{\mathfrak{k}}

\DeclareMathOperator{\tr}{Tr}

\DeclareMathOperator{\ad}{ad}

\newcommand{\id}{\mathds{1}}

\newcommand{\Y}{\mathrm{Y}}

\renewcommand{\phi}{\varphi}

\usepackage[symbol]{footmisc}

\renewcommand{\thefootnote}{\fnsymbol{footnote}}

\begin{document}

\begin{titlepage}

\begin{flushright}
\end{flushright}

\vskip 1.5in

\begin{center}
{\bf\Large{The Geroch Group in One Dimension}}

\vskip 0.5cm {Robert F. Penna\footnote[1]{rpenna@ias.edu} } 
\vskip 0.05in {\small{ \textit{Institute for Advanced Study}
\vskip -.4cm
{\textit{Einstein Drive, Princeton, NJ 08540 USA}}}
}
\end{center}
\vskip 0.5in
\baselineskip 16pt

\begin{abstract}  

We study the dimensional reduction of general relativity to a single null spacetime dimension.  The dimensionally reduced theory is a theory of six scalar fields governed by three constraints.  It has an infinite dimensional symmetry which is an enhanced version of the Geroch group.  To get a local action of the symmetry on solution space, we need to introduce an infinite tower of new fields and new constraints.  The symmetry appears to be a hyperbolic Kac-Moody algebra, with the caveat that some of the defining relations of the hyperbolic Kac-Moody algebra  are only checked ``order by order'' on the infinite tower of new fields.  This is a very mysterious Lie algebra with no known geometrical interpretation. It is not even clear how to enumerate a basis.  We explore this problem using the action of the algebra on solution space and find an intriguing connection to the representation theory of the symmetric group.  The symmetry described here might be related to the dynamics of gravity near spacelike singularities.

\end{abstract}


\end{titlepage}

\renewcommand*{\thefootnote}{\arabic{footnote}}
\setcounter{footnote}{0}

\tableofcontents

\section{Introduction}
\label{sec:intro}

The dimensional reduction of general relativity to two spacetime dimensions is an integrable sigma model with an infinite dimensional hidden symmetry called the Geroch group \cite{geroch1971method,geroch1972method,breitenlohner1987geroch,nicolai1991two}.  
Under the action of the Geroch group, the Minkowski metric can be mapped to any vacuum metric with two commuting Killing vectors.   This reduces the problem of deriving vacuum metrics with two commuting Killing vectors to pure algebra 
(for examples, see \cite{nicolai1991two,maison1999duality,katsimpouri2013inverse,penna2021einstein}).  
The existence of the Geroch group thus at least partially explains why general relativity, despite its complexity, admits many exact solutions \cite{griffiths2009exact,stephani2009exact}.  
There is a similar story for supergravity \cite{nicolai1987integrability,nicolai1989structure,penna2021lax}.

We have two motivations for asking how this story changes in one spacetime dimension.  The first is that the dynamics of general relativity becomes effectively one dimensional near spacelike singularities \cite{belinskii1970oscillatory,belinski2017cosmological}.  There is evidence for an emergent hyperbolic Kac-Moody symmetry in the near-singularity limit \cite{belinski2017cosmological,damour200210,damour2003cosmological}. This is a larger symmetry than the Geroch group.  The dynamics near the singularity is a function of a single timelike coordinate.  In the present work, we are going to study metrics that depend on a single null coordinate.  This is a simpler problem because some of the equations of motion are trivially satisfied.  We hope general features of our results will be useful for understanding the timelike case.

Our second motivation is that the hyperbolic Kac-Moody symmetry that seems to appear in one dimension is a mysterious and interesting mathematical structure in its own right \cite{feingold1983hyperbolic,borcherds1986vertex,kac1990infinite,carbone2014dimensions}.   It is an infinite dimensional Lie algebra with no known geometrical interpretation.  It is not even clear how to enumerate a basis.  In the last part of this work, we will try to shed some light on the structure of this algebra and we will find an intriguing connection to the representation theory of the symmetric group.

Nicolai \cite{nicolai1992hyperbolic} studied the dimensional reduction of supergravity to a single null dimension and argued for the emergence of a hyperbolic Kac-Moody symmetry in that case.   In that work, it was claimed that the fermions played an essential role and that the hyperbolic Kac-Moody symmetry would be trivial in pure general relativity.  One of the main messages of the present work is that the hyperbolic Kac-Moody symmetry acts even in pure general relativity.

At first glance, the dimensionally reduced theory looks almost trivial.  It is a theory of six scalar fields governed by three constraints.  To get a local action of the symmetry, it is necessary to introduce an infinite tower of new scalar fields (``dual potentials'') and new constraints.  
The symmetry is infinite dimensional but it has nine generators.  We obtain closed form expressions for the action of eight of the nine generators on solution space. The action of the ninth generator (called $f_{-1}$) is given ``order by order'' on the infinite tower of dual potentials.   We believe the action of $f_{-1}$ admits a unique extension to the whole tower of dual potentials but we do not have a proof.  It might be possible to prove the existence and uniqueness of $f_{-1}$ by carefully studying the constraints imposed by hyperbolic Kac-Moody symmetry, but we leave this interesting open problem for the future.

The commutators of the nine symmetry generators satisfy the defining relations of a hyperbolic Kac-Moody algebra, with the following caveats.  We have only checked the commutators involving $f_{-1}$ on the first few dual potentials in the infinite tower.  We verify most of the other commutators exactly on the whole infinite tower of dual potentials.   The exception is a pair of quadrilinear relations (equations \ref{eq:ffff1} and \ref{eq:ffff0}) which we only check on the first few dual potentials.  It should be possible to prove these relations exactly on the whole infinite tower using the fact that they only involve an affine Kac-Moody subalgebra of the full algebra, but this is another interesting problem for the future.

In the last part of this work, we try to shed some light on the structure of the symmetry algebra and we find an intriguing connection to the representation theory of the symmetric group.  The full symmetry algebra is generated by multiple commutators of  nine generators.
We consider subspaces generated by multiple commutators of three of the generators, $e_1$, $e_0$, and $e_{-1}$.  
We define\footnote{These are imaginary root spaces \cite{kac1990infinite}.  In the decomposition of Feingold  and Frenkel \cite{feingold1983hyperbolic}, they are level 1 imaginary roots.  The dimensions of these root spaces are known (equation \ref{eq:dimVintro}).   The dimensions of the other root spaces are not known in general \cite{carbone2014dimensions}.} $V_n$ to be the vector space generated by multiple commutators for which $e_1$ appears $n$ times, $e_0$ appears $n$ times, and $e_{-1}$ appears once.  The interpretation of the symmetry as a hyperbolic Kac-Moody algebra implies \cite{feingold1983hyperbolic}
\beq\label{eq:dimVintro}
\dim V_n = p(n) \,,
\eeq
where $p(n)$ is the number of partitions of $n$.  We verify this formula for $n=2,3,4,5$ by taking repeated commutators of the $e_i$ and finding linearly independent sets of basis vectors.  

To give an example, here is a basis for $V_3$:
\begin{align}
&\delta_{\eta_{0101}} + \dots \,, \label{eq:v1} \\
&\delta_{\eta_{0110}} + \delta_{\eta_{1010}} \dots \,, \label{eq:v2} \\
&\delta_{\eta_{1001}} + \dots \,. \label{eq:v3}
\end{align} 
The $\eta_I$ are dual potentials. The dots indicate ``subleading terms,'' which are terms involving higher order dual potentials.   Equations \eqref{eq:v1}--\eqref{eq:v3} describe the action of the $V_3$ basis vectors on solution space.  We have chosen a standard basis  for which the matrix of basis vectors is in reduced row echelon form.  

Now we observe that the number of terms in \eqref{eq:v1}--\eqref{eq:v3} are 
\beq
1, 2, 1
\eeq
and these are the dimensions of the irreducible representations of $S_3$.  Furthermore, there is a simple rule for generating    \eqref{eq:v1}--\eqref{eq:v3} in terms of the ``fundamental'' basis vector \eqref{eq:v2}.   This aligns well with the fact that the three irreducible representations of $S_3$ are exterior powers of the two dimensional fundamental representation.

We compute basis vectors of $V_n$ for $n=2,3,4,5$ and in each case we find a similar relationship to the representation theory of the symmetric group, $S_n$.  Recall that the number of irreducible representations of $S_n$ is $p(n) = \dim V_n$.  In general, we need to distinguish between irreducible representations of $S_n$ that are exterior powers of the fundamental representation, and irreducible representations of $S_n$ that are not exterior powers of the fundamental representation.  For the first kind of representation, we find basis vectors of $V_n$ with ``lengths'' matching the dimensions of the representations, and there is a simple rule for generating the leading parts of these basis vectors by taking ``powers'' of a ``fundamental'' basis vector.  Unfortunately, we do not know how to relate the other representations to the other basis vectors.

It is interesting to ask if the dimensional reduction of general relativity to one null dimension is an integrable system.  Roughly speaking, this should mean that the number of symmetries is equal to the number of degrees of freedom.  However, it is not clear how to make this definition precise.  In many cases, the right definition of integrability is the existence of a Lax operator with a spectral parameter \cite{lax1968integrals,woodhouse1988geroch,penna2021twistor}.  This definition makes sense when the hidden symmetry is an affine Kac-Moody algebra.  In that case, the symmetry algebra is (a central extension of) a loop algebra, and the spectral parameter is related to the loop parameter.  However, it is not clear what the right replacement for the Lax operator is when the hidden symmetry is a hyperbolic Kac-Moody algebra.  The problem of finding the right definition of integrability in this case appears to be bound up with the problem of finding a geometrical interpretation of hyperbolic Kac-Moody algebras.

\section{Solution Space}
\label{sec:solutions}

It is convenient for our purpose to regard the tetrad, $e^\alpha_\mu$, as the basic field of general relativity.  
Following Nicolai, we parametrize  $e^\alpha_\mu$ as
\beq\label{eq:tetrad}
e^a_\mu =
\begin{pmatrix}
\Delta^{-1/2}	&	0					&	0				&	0 \\
0			&	\Delta^{-1/2} \lambda		&	\Delta^{-1/2} \rho A	&	\Delta^{1/2} B_-  \\
0			&	0					&	\Delta^{-1/2} \rho 	&	\Delta^{1/2} B_2 \\
0			&	0					&	0				&	\Delta^{1/2}
\end{pmatrix}  .
\eeq
The tetrad depends on six functions, $\Delta, B_2, B_-, \rho, A$ and $\lambda$.  
(Four functions have been eliminated using a diffeomorpism.)  
For dimensional reduction, we assume these six functions are functions of a single spacetime coordinate: $\Delta=\Delta(u)$, $B_2=B_2(u),\dots\, .$

The metric is $g_{\mu \nu} = e_\mu^a e_\nu^b \eta_{ab}$, where $\eta_{ab}$ is the Minkowski metric, 
\beq
\eta_{ab} = 
\begin{pmatrix}
0	&	-1/2	&	0	&	0	\\
-1/2	&	0	&	0	&	0	\\
0	&	0	&	1	&	0	\\
0	&	0	&	0	&	1	
\end{pmatrix} .
\eeq
The coordinates are $(u,v,x^2,x^3)$.  Note that $u$ is  a null coordinate.

Einstein's equations, $G_{\mu\nu}=0$, are equivalent to three constraints:
\begin{align}
\lambda(u)^{-1} \rho(u)^{-1}(\lambda'(u)\rho'(u)-\lambda(u) \rho''(u))
		&= \frac{1}{2} \Delta(u)^{-2} \left( \Delta'(u)^2 + \frac{\Delta(u)^4B_2'(u)^2}{\rho(u)^2} \right) , \label{eq:constraint1}\\
B_-'(u) 	&= A(u) B_2'(u)\,, \label{eq:constraint2}\\
A'(u) 		&= 0 \,. \label{eq:constraint3}
\end{align}

So solution space is parameterized by six functions, ($\Delta(u)$, $B_2(u)$, $B_-(u)$, $\rho(u)$, $A(u)$, $\lambda(u)$), subject to three constraints (\ref{eq:constraint1}--\ref{eq:constraint3}).

There is fairly obvious $\SL(3,\RR)$ symmetry that maps solutions to solutions.  This is called the Matzner-Misner group and we will describe its action on solution space in the next section. 

To find more symmetries, we use the following trick.  First, we enlarge solution space by adding an additional function, $B=B(u)$, and an additional constraint,
\beq\label{eq:constraint4}
B' = \frac{\Delta^2}{\rho}  B_2' \,.
\eeq
Then we rewrite the constraint \eqref{eq:constraint1} in the simpler form
\beq \label{eq:constraint1b}
\lambda^{-1} \rho^{-1}(\lambda' \rho' - \lambda \rho'') = \frac{1}{2} \Delta^{-2} \left( \Delta'^2 + B'^2 \right) .
\eeq
Now the right hand side is suggestive of a hidden $\SL(2,\RR)$ symmetry.  To make this explicit, define
\beq
\Y^1 = 
\begin{pmatrix}
1	&	0	\\
0	&	-1
\end{pmatrix} , \quad
\Y^2 = 
\begin{pmatrix}
0	&	1	\\
1	&	0
\end{pmatrix} , \quad
\Y^3 = 
\begin{pmatrix}
0	&	1	\\
-1	&	0
\end{pmatrix} .
\eeq
This is a basis for $\gg = \sl(2,\RR)$. Decompose 
\beq
\gg = \kk \oplus \hh \,,
\eeq
where $\hh = \so(2)$ and $\kk$ is the orthogonal complement of $\hh$ in $\gg$.  Note that $\Y^1$ and $\Y^2$ are a basis for $\kk$ and $\Y^3$ is a basis for $\hh$.  Further define
\beq\label{eq:V}
V = 
\begin{pmatrix}
\Delta^{1/2}	&	B \Delta^{-1/2}	\\
0			&	\Delta^{-1/2}	
\end{pmatrix} \in \SL(2,\RR) \,,
\eeq
and decompose
\beq
V^{-1} V' = P + Q \,,
\eeq
where $P \in \kk$ and $Q \in \hh$.  This gives
\beq
P = \frac{1}{2}\Delta^{-1} ( \Delta' \Y^1 + B' \Y^2 ) \,, \quad
Q = \frac{1}{2}\Delta^{-1} B' \Y^3 \,,
\eeq
and
\beq
\tr(P^2) = \frac{1}{2}\Delta^{-2}(\Delta'^2 + B'^2) \,.
\eeq
Now the constraint \eqref{eq:constraint1b} becomes
\beq\label{eq:constraint1c}
\lambda^{-1} \rho^{-1}(\lambda' \rho' - \lambda \rho'') = \tr(P^2) \,.
\eeq
This form of the constraint suggests there is an action of $\SL(2,\RR)$  on $V$ that preserves the constraints.  This is called the Ehlers group and we will describe its action on solution space in the next section.  The Ehlers group and the Matzner-Misner group do not commute.  Together they generate an infinite dimensional symmetry, which is the Geroch group in one dimension.

Before describing this further, we need to confront the following problem.  The action of the Ehlers group on $B_2$ and $B_-$ is nonlocal (by \ref{eq:constraint2} and \ref{eq:constraint4}), and the action of the Matzner-Misner group on $B$ is nonlocal (by \ref{eq:constraint4}).  By ``nonlocal'' we mean that the infinitesimal variations of these fields involve integrals over $x$.   

To get local actions, we use a variant of our earlier trick: we enlarge solution space by adding additional functions and additional constraints.  In the present version of the trick, we need to add an infinite number of new functions and an infinite number of new constraints (one new constraint for each new function).  The new functions are described below.  We defer discussion of the new constraints to Section \ref{sec:newconstraints}.

The new functions (``dual potentials'') can be organized into two infinite pyramids:
\begin{align}
&\thinspace\phi 			\notag \\
\phi_0	&\quad	\phi_1 	\notag \\
\phi_{00}	\quad	\phi_{01}	&\quad	\phi_{10}	\quad \phi_{11} 		\notag \\
&\negthinspace\dots \,, 		\label{eq:tower1} \\
&\quad					\notag \\
&\thinspace\eta 			\notag \\
\eta_0	&\quad	\eta_1 	\notag \\
\eta_{00}	\quad	\eta_{01}	&\quad	\eta_{10}	\quad \eta_{11} 		\notag \\
&\negthinspace\dots \,.		\label{eq:tower2}
\end{align}
These functions are all functions of $x$: $\phi=\phi(u), \phi_0=\phi_0(u), \dots \, .$   Each of the pyramids has $2^n$ functions on the $n$th floor (floors are labeled from the top down, with $n=0,1,2,\dots$).  On the $n$th floor, the subscripts run over all possible strings of $0$'s and $1$'s of length $n$.

In the next section, we will describe the infinitesimal actions of the Matzner-Misner and Ehlers groups on solution space.

\section{Symmetry Algebra}
\label{sec:algebra}

In the first part of this section, we describe the infinitesimal actions of the Matzner-Misner and Ehlers groups on solution space.  
Along the way, in section \ref{sec:newconstraints}, we introduce the infinite tower of constraints on the infinite tower of dual potentials \eqref{eq:tower1}--\eqref{eq:tower2}.

\subsection{Matzner-Misner Algebra}

Consider the Matzner-Misner algebra first.  Recall (equation \ref{eq:tetrad})
\beq
e =
\begin{pmatrix}\label{eq:tetradb}
\Delta^{-1/2}	&	0					&	0				&	0 \\
0			&	\Delta^{-1/2} \lambda		&	\Delta^{-1/2} \rho A	&	\Delta^{1/2} B_-  \\
0			&	0					&	\Delta^{-1/2} \rho 	&	\Delta^{1/2} B_2 \\
0			&	0					&	0				&	\Delta^{1/2}
\end{pmatrix}  \,.
\eeq
Let
\begin{align}
E_0 &= 
\begin{pmatrix}
0	&	0	&	0	&	0	\\
0	&	0	&	0	&	0	\\
0	&	0	&	0	&	1	\\
0	&	0	&	0	&	0	\\
\end{pmatrix} \,, \quad
H_0 = 
\begin{pmatrix}
0	&	0	&	0	&	0	\\
0	&	0	&	0	&	0	\\
0	&	0	&	1	&	0	\\
0	&	0	&	0	&	-1	\\
\end{pmatrix} \,, \quad
F_0 = 
\begin{pmatrix}
0	&	0	&	0	&	0	\\
0	&	0	&	0	&	0	\\
0	&	0	&	0	&	0	\\
0	&	0	&	1	&	0	\\
\end{pmatrix} \,, 						\notag \\
E_{-1} &= 
\begin{pmatrix}
0	&	0	&	0	&	0	\\
0	&	0	&	1	&	0	\\
0	&	0	&	0	&	0	\\
0	&	0	&	0	&	0	\\
\end{pmatrix} \,, \quad
H_{-1} = 
\begin{pmatrix}
0	&	0	&	0	&	0	\\
0	&	1	&	0	&	0	\\
0	&	0	&	-1	&	0	\\
0	&	0	&	0	&	0	\\
\end{pmatrix} \,, \quad
F_{-1} = 
\begin{pmatrix}
0	&	0	&	0	&	0	\\
0	&	0	&	0	&	0	\\
0	&	1	&	0	&	0	\\
0	&	0	&	0	&	0	\\
\end{pmatrix} \,.			\label{eq:sl3}
\end{align}
These matrices generate $\sl(3,\RR)$.    Let $g$ be one of these matrices and consider the action on $e$ given by
\beq\label{eq:eaction}
e \rightarrow - ge + e h(g) \,,
\eeq
where $h(g)$ is a Lorentz transformation. We choose $h(g)$ to restore the upper triangular gauge of equation \eqref{eq:tetradb}.

It straightforward to work out the action of these matrices \eqref{eq:sl3} on the six basic fields ($\Delta$, $B_2$, $B_-$, $\rho$, $A$, $\lambda$) using equation \eqref{eq:eaction}.  This defines the infinitesimal action of the Matzner-Misner group on these fields\footnote{We differ from Nicolai \cite{nicolai1992hyperbolic} by factors of $2$ in $h_0(\lambda) = 2\lambda$ and $f_0(\lambda)  = -2B_2 \lambda$.}:
\begin{align}
e_0	=	& - \delta_{B_2}	+ \dots 	\,, \label{eq:e0a} \\
\quad \notag \\
h_0	=	& 2\Delta \delta_\Delta - 2B_2 \delta_{B_2} - B_- \delta_{B_-} 
			+ A \delta_A + 2 \lambda \delta_\lambda + \dots		\,,  \label{eq:h0a} \\
\quad \notag \\
f_0	=	& - 2\Delta B_2 \delta_{\Delta} +(B_2^2 - \rho^2 \Delta^{-2}) \delta_{B_2} 
			+ (B_- B_2 - \rho^2 \Delta^{-2} A) \delta_{B_-} 
			+ (B_- - B_2 A) \delta_A  \notag \\
			&  - 2 B_2 \lambda \delta_\lambda + \dots 		\label{eq:f0a} \,, \\
\quad \notag \\
e_{-1}	=	& - B_2 \delta_{B_-} - \delta_A + \dots	\,, \label{eq:ema} \\
\quad \notag \\
h_{-1}	=	& B_2 \delta_{B_2} - B_- \delta_{B_-} + \rho \delta_\rho  
				- 2 A \delta_A - \lambda \delta_\lambda 	+ \dots	\,, \label{eq:hma} \\
\quad \notag \\
f_{-1}	= 	& - B_- \delta_{B_2} - A \rho \delta_\rho + A^2 \delta_A + A \lambda \delta_\lambda	
			  + \dots \label{eq:fma} \,.
\end{align}
The dots $\dots$ are here because we still need to fix the action of these generators on $B$ and on the dual potentials \eqref{eq:tower1}--\eqref{eq:tower2}.

To fix the action of these generators on $B$, we make the assignments
\begin{align}
e_0(B)	&= 0 \,, \quad
h_0(B)	= 2B \,, \quad
f_0(B)	= -(\phi+2B_2 B + \rho) \,, \notag \\
e_{-1}(B)	&= 0 \,, \quad
h_{-1}(B)	= 0 \,, \quad
f_{-1}(B)	= 0 \,.
\end{align}
These are the simplest assignments that are compatible with the constraint equation \eqref{eq:constraint4} for $B$.  
The action of the Matzner-Misner algebra on the dual potentials \eqref{eq:tower1}--\eqref{eq:tower2} is described in subsections \ref{sec:toweractions} and \ref{sec:fm}.  First, we describe the infinitesimal action of the Ehlers group on the basic fields.

\subsection{Ehlers Algebra}

Earlier we introduced an $\SL(2,\RR)$-valued field (equation \ref{eq:V}),
\beq\label{eq:Va}
V = 
\begin{pmatrix}
\Delta^{1/2}	&	B \Delta^{-1/2}	\\
0			&	\Delta^{-1/2}	
\end{pmatrix} .
\eeq
Define
\beq
E_1 = 
\begin{pmatrix}
0	&	1	\\
0	&	0	\\
\end{pmatrix} , \quad
H_1 = 
\begin{pmatrix}
1	&	0	\\
0	&	-1	\\
\end{pmatrix} , \quad
F_1 = 
\begin{pmatrix}
0	&	0	\\
1	&	0	\\
\end{pmatrix} . \label{eq:sl2}
\eeq
These matrices generate $\sl(2,\RR)$.  Let $g$ be one of these matrices and consider the action on $V$ given by
\beq\label{eq:Vaction}
V \rightarrow -g V + Vh(g) \,,
\eeq
where $h(g)$ is a compensating $\so(2)$ transformation. We choose $h(g)$ to restore the upper triangular gauge of equation \eqref{eq:Va}.

It is straightforward to work out the action of the Ehlers algebra on $\Delta$ and $B$ using equation \eqref{eq:Vaction}.
We extend this action to the other basic fields by making the simplest possible assignments that are compatible with the constraint equations \eqref{eq:constraint1}--\eqref{eq:constraint4}.  This gives the action of the Ehlers algebra on the basic fields ($\Delta$, $B_2$, $B_-$, $\rho$, $A$, $\lambda$, $B$):
\begin{align}
e_1	=	& - \delta_B + \dots  \,, \label{eq:e1a}\\
\quad \notag \\
h_1	= 	& - 2\Delta \delta_\Delta + 2B_2 \delta_{B_2} + 2 B_- \delta_{B_-} - 2B \delta_B + \dots\,, \label{eq:h1a} \\
\quad \notag \\
f_1	= 	& 2 \Delta B \delta_\Delta + \phi \delta_{B_2} + \eta \delta_{B_-} + (B^2 - \Delta^2) \delta_B 
		 + \dots	\,.  \label{eq:f1a}
\end{align}
The dots $\dots$ are here because we still need to fix the action of these generators on the dual potentials \eqref{eq:tower1}--\eqref{eq:tower2}.  This is the subject of the next subsection.

\subsection{The Action of the Generators on $\phi_I$ and $\eta_I$}
\label{sec:toweractions}

The following notation will be useful for manipulating the dual potentials \eqref{eq:tower1}--\eqref{eq:tower2}.  Let $I$ be an index that runs over all possible strings of $0$'s and $1$'s (including the empty string).  For example, $\phi_I$ can stand for any function in  the first list \eqref{eq:tower1}.     Let $n_0^I$ be the number of zeros in $I$ and let $n_1^I$ be the number of ones in $I$.  

We extend the actions of $f_1$ and $f_0$ to $\phi_I$ and $\eta_I$ by defining
\begin{align}
f_1	= 	& 2 \Delta B \delta_\Delta + \phi \delta_{B_2} + \eta \delta_{B_-} + (B^2 - \Delta^2) \delta_B 
		 + \phi_{I1} \delta_{\phi_I} + \eta_{I1} \delta_{\eta_I}	\,,  \label{eq:f1b} \\
\quad \notag \\
f_0	=	& - 2\Delta B_2 \delta_{\Delta} +(B_2^2 - \rho^2 \Delta^{-2}) \delta_{B_2} 
			+ (B_- B_2 - \rho^2 \Delta^{-2} A) \delta_{B_-} 
			+ (B_- - B_2 A) \delta_A  \notag \\
			&  - 2 B_2 \lambda \delta_\lambda -(\phi+2B_2B +2\rho) \delta_B
			 + \phi_{I0} \delta_{\phi_I} + \eta_{I0} \delta_{\eta_I}		\label{eq:f0b} \,.
\end{align}
Terms with repeated $I$ indices are sums over all $I$. 
With these definitions,
\beq\label{eq:potentials}
\phi = f_1(B_2) \,, \quad
\eta = f_1(B_-) \,,
\eeq
and
\begin{align}
\phi_{I0} &= f_0(\phi_I) \,, \quad
\eta_{I0} = f_0(\eta_I) \,, \label{eq:add0}\\
\phi_{I1} &= f_1(\phi_I) \,, \quad
\eta_{I1} = f_1(\eta_I) \,. \label{eq:add1}
\end{align}
So we may write ($i_j=0$ or $1$)
\begin{align}
\phi_I &= \phi_{i_1\dots i_n} = f_{i_n} \cdots f_{i_1} \phi \,, \\
\eta_I &= \eta_{i_1\dots i_n} = f_{i_n} \cdots f_{i_1} \eta \,.
\end{align}
The dual potentials enter here because the actions of $f_1$ on $B_2$ on $B_-$ are nonlocal when expressed in terms of the original fields (by \ref{eq:constraint2} and \ref{eq:constraint4}).

Now consider the problem of defining the action of the $h_i$ ($i=-1,0,1$) on  $\phi_I$ and $\eta_I$.  
Let 
\beq
A_{ij} = 
\begin{pmatrix}
2	&	-2	&	0	\\
-2	&	2	&	-1	\\
0	&	-1	&	2
\end{pmatrix} .
\eeq
Observe that if $X$ is one of the basic fields, ($\Delta$, $B_2$, $B_-$, $\rho$, $A$, $\lambda$, $B$), then
\beq
[h_i, f_j]X = - A_{ij} f_j(u) \label{eq:hcom} \,.
\eeq
The commutators are vector field commutators:
\beq
[h_i,f_j] X \equiv h_i(f_j(u)) - f_j(h_i(u)) \,.
\eeq 
We need to define
\beq\label{eq:h1phiI}
h_1(\phi_I) = h_1 f_{i_n} \cdots f_{i_1} \phi \,.
\eeq
If we move $h_1$ to the right using \eqref{eq:hcom}, we obtain
$h_1(\phi_I) = - 2 ( n_1^I - n_0^I ) \phi_I.$
We use this equation to define $h_1(\phi_I)$.  The other coefficients are defined similarly.  

We thus obtain
\begin{align}
h_1	= 	& - 2\Delta \delta_\Delta + 2B_2 \delta_{B_2} + 2 B_- \delta_{B_-} - 2B \delta_B
			- 2 ( n_1^I - n_0^I ) \phi_I \delta_{\phi_I} 
			- 2 ( n_1^I - n_0^I ) \eta_I \delta_{\eta_I} 		\,, \label{eq:h1b} \\
\quad \notag \\
h_0	=	& 2\Delta \delta_\Delta - 2B_2 \delta_{B_2} - B_- \delta_{B_-} 
			+ A \delta_A + 2 \lambda \delta_\lambda + 2B \delta_{B}	\notag \\
		& + 2 ( n_1^I - n_0^I ) \phi_I \delta_{\phi_I} 
			+ (2 (n_1^I - n_0^I) + 1 ) \eta_I \delta_{\eta_I} 		\,,  \label{eq:h0b} \\
\quad \notag \\
h_{-1}	=	& B_2 \delta_{B_2} - B_- \delta_{B_-} + \rho \delta_\rho  
				- 2 A \delta_A - \lambda \delta_\lambda
				+ (n_0^I + 1 ) \phi_I \delta_{\phi_I} 
				+  (n_0^I - 1 ) \eta_I \delta_{\eta_I}		\,. \label{eq:hmb}  
\end{align}
These equations define the actions of $h_1$, $h_0$, and $h_{-1}$ on solution space.

The actions of the $e_i$ on  the dual potentials \eqref{eq:tower1}--\eqref{eq:tower2} are defined similarly.  
We note that if $X$ is one of the basic fields, ($\Delta$, $B_2$, $B_-$, $\rho$, $A$, $\lambda$, $B$), then
\beq\label{eq:ecom}
[e_i, f_j ] X= \delta_{ij}X \,.
\eeq
To define
\beq
e_1(\phi_I) = e_1 f_{i_n} \cdots f_{i_1} \phi ,
\eeq
we move $e_1$ to the right using \eqref{eq:ecom}.  This gives
\begin{align}
e_1	=	& - \delta_B 
		 + e_1^{\phi_I} \delta_{\phi_I} + e_1^{\eta_I} \delta_{\eta_I}  \,, \label{eq:e1b}\\
\quad \notag \\
e_0	=	& - \delta_{B_2}	
		  + e_0^{\phi_I} \delta_{\phi_I} + e_0^{\eta_I} \delta_{\eta_I} 	\,, \label{eq:e0b} \\
\quad \notag \\
e_{-1}	=	& - B_2 \delta_{B_-} - \delta_A 	
			 -\phi_I \delta_{\eta_I}		\,, \label{eq:emb} 
\end{align}
with
\begin{align}
e_1^{\phi_I}	&=  \sum_{j=1}^n f_{i_n} \cdots h_1 \delta_{i_j}^1 \cdots f_{i_1} \phi
				+ 2 f_{i_n} \cdots f_{i_1} B_2  \,, \label{eq:e1coef1}\\
e_1^{\eta_I}	&= \sum_{j=1}^n f_{i_n} \cdots h_1 \delta_{i_j}^1 \cdots f_{i_1} \eta 
				+ 2 f_{i_n} \cdots f_{i_1} B_- \,,  \label{eq:e1coef3}
\end{align}
and
\begin{align}
e_0^{\phi_I}	&= \sum_{j=1}^n f_{i_n} \cdots h_0 \delta_{i_j}^0 \cdots f_{i_1} \phi \,, \label{eq:e0coef1}\\
e_0^{\eta_I}	&= \sum_{j=1}^n f_{i_n} \cdots h_0 \delta_{i_j}^0 \cdots f_{i_1} \eta \,.  \label{eq:e0coef3}
\end{align}
$\delta_{i_j}^0$ and $\delta_{i_j}^1$ are Kronecker delta functions.  

All that remains is to define the action of $f_{-1}$ on $\phi_I$ and $\eta_I$.  This step presents special difficulties and we will return to it in subsection \ref{sec:fm}.  First, we introduce the constraint equations for $\phi_I$ and $\eta_I$.

\subsection{Constraint Equations for $\phi_I$ and $\eta_I$}
\label{sec:newconstraints}

Write the constraints \eqref{eq:constraint2} and \eqref{eq:constraint4} as
\begin{align}
B_2'	&= \rho^{-1} \Delta^2  B' \,, \\
B_-' 	&= A(u) B_2' \,.
\end{align}
Act on these constraints with $f_1$ to obtain (use \ref{eq:f1b}--\ref{eq:potentials})
\begin{align}
\phi'	&=  -2 B B_2' - 2\rho \Delta^{-1} \Delta' \,, \label{eq:phiconstraint}\\
\eta'	&= -2 A B B_2' - 2\rho A \Delta^{-1} \Delta' \,. \label{eq:etaconstraint}
\end{align}
These are the constraint equations for $\phi$ and $\eta$.  The other constraint equations are defined similarly.  For example, to obtain the constraint equation for $\phi_I = f_{i_n} \cdots f_{i_1} \phi$ , act on \eqref{eq:phiconstraint} with $f_{i_n} \cdots f_{i_1}$ ($i_j=0$ or $1$).

\subsection{The Action of $f_{-1}$ on $\phi_I$ and $\eta_I$}
\label{sec:fm}

Now return to the problem of defining the action of $f_{-1}$ on $\phi_I$ and $\eta_I$.  This problem is complicated by the fact that we do not have a closed form expression\footnote{Nicolai notes that $[f_0,f_{-1}](B_2) = (A f_0 + (B_- - A B_2))(B_2)$, but this relation is not true on all the fields.  For example, it is not true on  $\Delta$.  We have not been able to find a closed form expression for $[f_0,f_{-1}]$ that is valid on all the fields.} for $[f_{-1},f_0]$.

Acting with $f_{-1}$ on the constraint equations \eqref{eq:phiconstraint}--\eqref{eq:etaconstraint} for $\phi$ and $\eta$ gives
\begin{align}
\partial_x (f_{-1}(\phi)) &= -\partial_x \eta \,, \\
\partial_x (f_{-1}(\eta)) &= 0 \,.
\end{align}
Integrating and setting the integration constants to zero gives
\begin{align}
f_{-1}(\phi) &= -\eta \,, \label{eq:fmphi}\\ 
f_{-1}(\eta) &= 0 \,. \label{eq:fmeta}
\end{align}
These equations define the action of $f_{-1}$ on $\phi$ and $\eta$.

Actually this discussion was somewhat ambiguous, because the action of $f_{-1}$ on the dual potentials is only defined up to total derivatives.  However, it it not too hard to see that equations \eqref{eq:fmphi}--\eqref{eq:fmeta} are uniquely fixed by the further requirement that the commutators of $f_{-1}$ with the other generators obey the defining relations of the hyperbolic Kac-Moody algebra discussed in the next section.

In fact, we have checked that asking for hyperbolic Kac-Moody symmetry uniquely fixes
\begin{align}
f_{-1}(\phi_1)	&=	-\eta_1 \,, \\
f_{-1}(\phi_0)	&=	- 2(\eta_0 + B_2 \eta  - B_- \phi) \,, \\
f_{-1}(\eta_1)	&=	0 \,.
\end{align}
We believe that asking for hyperbolic Kac-Moody symmetry uniquely fixes the action of $f_{-1}$ on the whole infinite tower of dual potentials but we do not have a proof.

\subsection{Summary}

In summary, the action of the symmetry generators on solution space is
\begin{align}
e_1	=	& - \delta_B 
		 + e_1^{\phi_I} \delta_{\phi_I} + e_1^{\eta_I} \delta_{\eta_I}  \,, \label{eq:e1}\\
\quad \notag \\
h_1	= 	& - 2\Delta \delta_\Delta + 2B_2 \delta_{B_2} + 2 B_- \delta_{B_-} - 2B \delta_B
			- 2 ( n_1^I - n_0^I ) \phi_I \delta_{\phi_I} 
			- 2 ( n_1^I - n_0^I ) \eta_I \delta_{\eta_I} 		\,, \label{eq:h1} \\
\quad \notag \\
f_1	= 	& 2 \Delta B \delta_\Delta + \phi \delta_{B_2} + \eta \delta_{B_-} + (B^2 - \Delta^2) \delta_B 
		 + \phi_{I1} \delta_{\phi_I} + \eta_{I1} \delta_{\eta_I}	\,,  \label{eq:f1} \\
\quad \notag \\
e_0	=	& - \delta_{B_2}	
		  + e_0^{\phi_I} \delta_{\phi_I} + e_0^{\eta_I} \delta_{\eta_I} 	\,, \label{eq:e0} \\
\quad \notag \\
h_0	=	& 2\Delta \delta_\Delta - 2B_2 \delta_{B_2} - B_- \delta_{B_-} 
			+ A \delta_A + 2 \lambda \delta_\lambda + 2B \delta_{B}	\notag \\
		& + 2 ( n_1^I - n_0^I ) \phi_I \delta_{\phi_I} 
			+ (2 (n_1^I - n_0^I) + 1 ) \eta_I \delta_{\eta_I} 		\,,  \label{eq:h0} \\
\quad \notag \\
f_0	=	& - 2\Delta B_2 \delta_{\Delta} +(B_2^2 - \rho^2 \Delta^{-2}) \delta_{B_2} 
			+ (B_- B_2 - \rho^2 \Delta^{-2} A) \delta_{B_-} 
			+ (B_- - B_2 A) \delta_A  \notag \\
			&  - 2 B_2 \lambda \delta_\lambda - (\phi + 2B_2B + 2\rho) \delta_B
			 + \phi_{I0} \delta_{\phi_I} + \eta_{I0} \delta_{\eta_I}		\label{eq:f0} \,, \\
\quad \notag \\
e_{-1}	=	& - B_2 \delta_{B_-} - \delta_A 	
			 -\phi_I \delta_{\eta_I}		\,, \label{eq:em} \\
\quad \notag \\
h_{-1}	=	& B_2 \delta_{B_2} - B_- \delta_{B_-} + \rho \delta_\rho  
				- 2 A \delta_A - \lambda \delta_\lambda
				+ (n_0^I + 1 ) \phi_I \delta_{\phi_I} 
				+  (n_0^I - 1 ) \eta_I \delta_{\eta_I}		\,, \label{eq:hm} \\
\quad \notag \\
f_{-1}	= 	& - B_- \delta_{B_2} - A \rho \delta_\rho + A^2 \delta_A + A \lambda \delta_\lambda	
			  + f_{-1}^{\phi_I} \delta_{\phi_I} + f_{-1}^{\eta_I} \delta_{\eta_I}  \label{eq:fm} \,.
\end{align}

These nine transformations \eqref{eq:e1}--\eqref{eq:fm} generate the Geroch algebra in one dimension.  The Geroch algebra in two dimensions is generated by $e_1$, $h_1$, $f_1$, $e_0$, $h_0$, and $f_0$ alone.  The Geroch algebra in two dimensions can be realized on the subspace of solution space with $B_- = A = \eta_I = 0$, but the enhanced symmetry in one dimension cannot be realized on this subspace because the enhanced symmetry mixes $B_-$, $A$ and $\eta_I$ with the other fields.

It is interesting to note that none of the coefficients in \eqref{eq:e1}--\eqref{eq:fm} involve spatial derivatives of the fields.  So we could fix $x$ and obtain an action of the algebra on an infinite dimensional space of real valued constants.

\section{Invariance of the Constraints}
\label{sec:constraints}

It is straightforward to check that the nine symmetry generators \eqref{eq:e1}--\eqref{eq:fm} map the four basic constraint equations \eqref{eq:constraint1}--\eqref{eq:constraint4},
\begin{align}
\lambda^{-1} \rho^{-1}(\lambda' \rho'-\lambda \rho'')
		&= \frac{1}{2} \Delta^{-2} ( \Delta'^2 + B'^2 ) \,, \\
B_-'	 	&= A B_2'\,, \\
A'		&= 0 \,, \\
B'		&= \frac{\Delta^2}{\rho}  B_2' \,,
\end{align}
to linear combinations of constraints.  So these four constraints are preserved.  

Now consider the constraint equations for $\phi_I$ and $\eta_I$.  Define
\begin{align}
C_{B_2}	&= B_2' - \rho \Delta^{-2} B' \,, \\
C_{B_-}	&= B_-' - A B_2' \,,
\end{align}
and
\begin{align}
C_{\phi_I} &= f_{i_n} \cdots f_{i_1} f_1 C_{B_2} \,, \label{eq:CphiI}\\
C_{\eta_I} &= f_{i_n} \cdots f_{i_1} f_1 C_{B_-} \label{eq:CetaI}\,.
\end{align}
The constraint equations for $\phi_I$ and $\eta_I$ are (section \ref{sec:newconstraints})
\beq
C_{\phi_I} =0 \,, \quad 
C_{\eta_I} = 0 \,.
\eeq
The proof that $f_1$ and $f_0$ preserve these constraints is trivial because ($j=0$ or $1$)
\beq
f_j C_{\phi_I} = C_{\phi_{Ij}} \,, \quad
f_j C_{\eta_I} = C_{\eta_{Ij}} \,.
\eeq

The definition of $f_{-1}$ implies 
\beq\label{eq:fmC}
f_{-1}C_{\phi_I} = f_{-1}C_{\eta_I} = 0 \,,
\eeq
and so $f_{-1}$ also preserves the constraints, with the caveat that the definition of $f_{-1}$ only makes sense if $f_{-1}C_{\phi_I}$,  and $f_{-1}C_{\eta_I}$ are total derivatives.  As noted in section \ref{sec:fm}, we have checked this assumption for the first few fields in the two infinite towers \eqref{eq:tower1}--\eqref{eq:tower2}, and we believe it holds for all the fields, but we do not have a proof.

The definitions of the generators \eqref{eq:e1}--\eqref{eq:fm} give
\beq
[h_i , f_j] = -A_{ij} f_j \,, \quad (j=0,1) \,.
\eeq
These commutation relations and \eqref{eq:CphiI}--\eqref{eq:CetaI} imply that $h_1$, $h_0$, and $h_{-1}$ map the constraint equations for $\phi_I$ and $\eta_I$ to linear combinations of constraints.

It is equally straightforward to check that 
\beq
[e_i , f_j ] = \delta_{ij} \,, \quad (j=0,1) \,.
\eeq
and so 
$e_1$, $e_0$, and $e_{-1}$ map the 
the constraint equations for $\phi_I$ and $\eta_I$ to linear combinations of constraints.

This completes the verification that the nine generators of the Geroch algebra \eqref{eq:e1}--\eqref{eq:fm} preserve the constraints (subject to the caveat below equation \ref{eq:fmC}).

\section{Hyperbolic Kac-Moody Commutators}
\label{sec:hyperbolic}

Recall
\beq
A_{ij} = 
\begin{pmatrix}
2	&	-2	&	0	\\
-2	&	2	&	-1	\\
0	&	-1	&	2
\end{pmatrix} .\label{eq:Aij}
\eeq
The  goal for the remainder of this section is to show that the generators \eqref{eq:e1}--\eqref{eq:fm} satisfy
\begin{align}
[ h_i , h_j ] &= 0 \,, \quad
[ h_i, e_j ] = A_{ij} e_j \,,  \label{eq:com1} \\
[ e_i, f_j ] &= \delta_{ij} h_j \,,\quad
[ h_i, f_j ] = -A_{ij} f_j \,, \label{eq:com2}
\end{align}
and for $i\neq j$,
\beq\label{eq:com3}
(\ad e_i)^{1-A_{ij} } (e_j ) = 0  \,, \quad
(\ad f_i)^{1-A_{ij} } (f_j ) = 0 \,.
\eeq
These equations \eqref{eq:Aij}--\eqref{eq:com3} define the hyperbolic Kac-Moody algebra with Cartan matrix $A_{ij}$.

It  is straightforward to verify the first set of relations \eqref{eq:com1} using the definitions of the generators \eqref{eq:e1}--\eqref{eq:fm}.  It is also straightforward to verify the relations \eqref{eq:com2} for $j=0,1$.  The relations \eqref{eq:com2} with $j=-1$ involve $f_{-1}$, and we do not have a closed form expression for $f_{-1}$, so these relations must be checked order by order on the infinite tower of dual potentials \eqref{eq:tower1}--\eqref{eq:tower2}.  We have checked these relations on the first few fields, and we believe they hold on the whole tower, but we do not have a proof.

It is not too hard to establish $(\ad e_i)^{1-A_{ij} } (e_j ) = 0$.  First, we check this relation directly on the seven basic fields ($\Delta$, $B_2$, $B_-$, $\rho$, $A$, $\lambda$, $B$).  Then we use \eqref{eq:com1}--\eqref{eq:com2} (for $j=0,1$) and the Jacobi identity to get
\beq
[(\ad e_i)^{1-A_{ij} } (e_j ), f_0 ]  = [(\ad e_i)^{1-A_{ij} } (e_j ), f_1] = 0 \,.
\eeq
It follows immediately that $(\ad e_i)^{1-A_{ij} } (e_j ) = 0$ on
\begin{align}
\phi_I	&= f_{i_n} \cdots f_{i_1} f_1 B_2	\,,	\\
\eta_I 	&= f_{i_n} \cdots f_{i_1} f_1 B_-	\,.
\end{align}
Thus $(\ad e_i)^{1-A_{ij} } (e_j ) = 0$ on all fields.

Finally consider
\beq
(\ad f_i)^{1-A_{ij} } (f_j ) = 0 \,.
\eeq
Unpacking this gives three equations involving $f_{-1}$,
\begin{align}
[ f_1, f_{-1} ]		&=0	\,,	\label{eq:ff}\\
[ f_0, [ f_0, f_{-1} ]	&=0	\,,	\label{eq:fff0}\\
[ f_{-1}, [ f_{-1}, f_0 ]	&=0	\,,	\label{eq:ffm}
\end{align}
and two equations not involving $f_{-1}$,
\begin{align}
[ f_1, [ f_1, [ f_1, f_0 ] ] ]	&= 0 \,, 	\label{eq:ffff1}\\
[ f_0, [ f_0, [ f_0, f_1 ] ] ]	&= 0 \,.	\label{eq:ffff0}
\end{align}
The first three equations \eqref{eq:ff}--\eqref{eq:ffm} need to be checked order by order on the infinite tower of dual potentials because we do not have closed form expressions for $f_{-1}$.  We have checked these equations on the basic fields ($\Delta$, $B_2$, $B_-$, $\rho$, $A$, $\lambda$, $B$) and on $\phi$ and $\eta$, and we expect they hold on all $\phi_I$ and $\eta_I$, but we do not have a proof.

To satisfy the quadrilinear relations \eqref{eq:ffff1} and \eqref{eq:ffff0}, we need to impose an infinite tower of new algebraic constraints on $\phi_I$ and $\eta_I$.   Here is the first one:
\beq
[ f_1, [ f_1, [ f_1, f_0 ] ] ](\Delta) = 4 \Delta \phi_{11} = 0 \,.
\eeq
So we need to impose $\phi_{11} = 0$ as a further constraint on solution space.  For this to be consistent, it must be compatible with our earlier constraint,
\beq
C_{\phi_{11}}  = f_1 f_1 f_1 (B_2' - \rho \Delta^{-2} B') = 0 \,.
\eeq
We find (using $C_{\phi_1} = C_\phi =0 $) that $C_{\phi_{11}} = 0$ reduces to $\phi_{11}'  = 0$, so these constraints are compatible.

We also need to check that the new constraint is preserved by the action of the generators \eqref{eq:e1}--\eqref{eq:fm}.  This is straightforward for $h_i$ and $e_i$.  

Repeated applications of $f_0$ and $f_1$ map the new constraint, $\phi_{11} = 0$, to an infinite tower of new constraints, $\phi_{11I} = 0$.  We will ignore these new constraints for now, because they are constraints on higher order dual potentials.

Finally, we compute $f_{-1} (\phi_{11}) = -\eta_{11}$.  This implies we need to impose the further constraint $\eta_{11} = 0$.  
It is not too hard to check that $\eta_{11} = 0$ is compatible with $C_{\eta_{11}} = 0$ and it is preserved by the $e_i$, $h_i$, and $f_{-1}$.  To have invariance under $f_1$ and $f_0$, we need to impose the higher level constraints $\eta_{11I} = 0$.

We believe the quadrilinear relations \eqref{eq:ffff1} and \eqref{eq:ffff0} can be consistently satisfied order by order, by imposing additional algebraic constraints on the dual potentials, but we do not have a proof.  It might be possible to prove this conjecture using the fact that these relations only involve an affine Kac-Moody subalgebra of the full algebra.  

This completes the verification that the commutators of the generators obey the defining relations of a hyperbolic Kac-Moody algebra, with the caveats that commutators involving $f_{-1}$ and the quadrilinear relations \eqref{eq:ffff1}--\eqref{eq:ffff0} have only been checked ``order by order'' on the first few fields in the infinite tower.

\section{Imaginary Roots}
\label{sec:roots}

In this final section, we would like to explore the structure of the symmetry algebra.  It is an infinite dimensional Lie algebra with nine generators.  It is not clear how to enumerate a basis.  

Let $V_n$ be the vector space generated by multiple commutators for which $e_1$ appears $n$ times, $e_0$ appears $n$ times, and $e_{-1}$ appears once.  These are imaginary root spaces \cite{kac1990infinite}.  In the decomposition of Feingold  and Frenkel \cite{feingold1983hyperbolic}, they are level 1 imaginary roots.  The dimensions of these root spaces are known \cite{feingold1983hyperbolic}:
\beq\label{eq:dimV}
\dim V_n = p(n) \,,
\eeq
where $p(n)$ is the number of partitions of $n$.  

The appearance of $p(n)$ suggests the structure of the algebra might be related to the representation theory of the symmetric group, $S_n$, because $p(n)$ is the number of irreducible representations of $S_n$.

We compute bases for the $V_n$ using equations \eqref{eq:e1}--\eqref{eq:fm} for the generators.   In each case, we choose a standard basis for which the matrix of basis vectors is in reduced row echelon form.  The results for $n=2,3,4,5$ are summarized in Table \ref{tab:Vn}.  We show the leading order terms of the basis vectors, by which we mean that we drop terms involving higher order dual potentials.  In each case,  $\dim V_n = p(n)$.

Define the ``lengths'' of the vectors in Table \ref{tab:Vn} to be the number of leading order terms.  It turns out that the lengths are related to dimensions of irreducible representations of the symmetric group.  To explain this, we need to distinguish between irreducible representations of the symmetric group that are exterior powers of the fundamental representation, and irreducible representations of the symmetric group that are not exterior powers of the fundamental representation.  The dimensions of the former kind of representations all appear in Table \ref{tab:Vn} as lengths of basis vectors.  

What is more, for each $n$, there is a simple rule for generating these basis vectors in terms of a ``fundamental'' basis vector.  
Define an index permutation operator, $\sigma_k$, by 
\beq
\sigma_k(\partial_{\eta_{i_1,j_1, \dots, i_k, j_k, \dots i_n,j_n}})
	= \partial_{\eta_{i_1,j_1, \dots, j_k, i_k, \dots i_n,j_n}} \,.
\eeq
In other words, $\sigma(k)$ permutes $i_k$ and $j_k$.

The black basis vectors in Table \ref{tab:Vn} can be represented as permutation operators acting on $\partial_{\eta_{0101\dots01}}$.  For example, the basis of $V_3$ is
\begin{align}
&\id \,, \label{eq:V31}\\
&\sigma(2) + \sigma(1)\sigma(2) \,, \label{eq:V32}\\
&\sigma(1) \,. \label{eq:V33}
\end{align}
Now we regard the middle entry as the ``fundamental'' basis vector (corresponding to the two dimensional fundamental representation of $S_3$), and observe that the other two vectors can be obtained by taking powers of \eqref{eq:V32}.

Unfortunately, we do not know how to relate irreducible representations of the symmetric group that are not exterior powers of the fundamental representation to the other basis vectors in Table \ref{tab:Vn}.  To understand this better, it would be interesting to relate our construction to the construction of Feingold and Frenkel \cite{feingold1983hyperbolic}.

\begin{table}[h]
\begin{center}
\begin{tabular}{ |c|c|c|c| } 
\hline
$n$	&	$p(n)$	&	$V_n$ basis vectors (leading terms only)	&	length	\\
 \hline
 	& 		&					&		\\
	&		&	$\delta_{\eta_{01}}$	&	1	\\
2	&	2	&					&		\\ 
	&		&	$\delta_{\eta_{10}}$	&	1	\\ 
	&		&					&		\\
\hline
 	& 		&										&		\\
	&		&	$\delta_{\eta_{0101}}$					&	1	\\ 
 	& 		&										&		\\
3	&	3	&	$\delta_{\eta_{0110}} + \delta_{\eta_{1010}}$	&	2	\\ 
 	& 		&										&		\\
	&		&	$\delta_{\eta_{1001}}$ 					&	1	\\ 
 	& 		&										&		\\
\hline
 	& 		&																&		\\
	&		&	$\delta_{\eta_{010101}}$											&	1	\\ 
 	& 		&																&		\\
	&		&	$\delta_{\eta_{010110}} + \delta_{\eta_{011010}} + \delta_{\eta_{101010}}$	&	3	\\ 
 	& 		&																&		\\
4	&	5	&	$\delta_{\eta_{011001}} + \delta_{\eta_{100101}} + \delta_{\eta_{101001}}$	&	3	\\ 
 	& 		&																&		\\
	&		&	$\delta_{\eta_{100110}}$											&	1	\\ 
 	& 		&																&		\\
	&		&	$\color{gray} \delta_{\eta_{010011}} + 3\delta_{\eta_{100011}} + \delta_{\eta_{100101}} $	&		\\ 
 	& 		&																&		\\
\hline
 	& 		&																						&		\\
	&		&	$\delta_{\eta_{01010101}}$																&	1	\\
 	& 		&																						&		\\
	&		&	$\delta_{\eta_{01010110}} + \delta_{\eta_{01011010}} + \delta_{\eta_{01101010}} + \delta_{\eta_{10101010}}$	&	4	\\
 	& 		&																						&		\\
	&		&	$\delta_{\eta_{01011001}} + \delta_{\eta_{01100101}} + \delta_{\eta_{01101001}} + \delta_{\eta_{10010101}} + \delta_{\eta_{10100101}} + \delta_{\eta_{10101001}}$ &	6	\\
 	& 		&																						&		\\
	&		&	$\delta_{\eta_{01100110}} + \delta_{\eta_{10010110}} + \delta_{\eta_{10011010}} + \delta_{\eta_{10100110}}$	&	4	\\
 	& 		&																						&		\\
5	&	7	&	$\delta_{\eta_{10011001}}$																&	1	\\
 	& 		&																						&		\\
	&		&	$\begin{array}{c}
		\color{gray}	\delta_{\eta_{01001101}}	+  3 \delta_{\eta_{01001110}} - \delta_{\eta_{0101101  0}} - \delta_{\eta_{01101010}} +  3 \delta_{\eta_{10001101}}	\\
		\color{gray}		+\thinspace  9 \delta_{\eta_{10001110}} + \delta_{\eta_{10010101}} +  3 \delta_{\eta_{10010110}} - \delta_{\eta_{10101010}}
		\end{array}$																					&		\\
 	& 		&																						&					\\
	&		&	$\begin{array}{c} \color{gray}
		\color{gray}	\delta_{\eta_{01001011}}	- 3 \delta_{\eta_{01001110}} +  2 \delta_{\eta_{01010011}} + \delta_{\eta_{01011010}} +  3 \delta_{\eta_{01100011}}	\\
		\color{gray}		+\thinspace \delta_{\eta_{01100101}} + \delta_{\eta_{01101010}} +  3 \delta_{\eta_{10001011}} -  9 \delta_{\eta_{10001110}}			\\
		\color{gray}		+\thinspace 4 \delta_{\eta_{10010011}} + \delta_{\eta_{10010101}} -  3 \delta_{\eta_{10010110}} +  3 \delta_{\eta_{10100011}}			\\
		\color{gray}		+\thinspace \delta_{\eta_{10100101}} + \delta_{\eta_{10101010}} 
		\end{array}$ 																					&		\\ 
 	& 		&																						&		\\
\hline
\end{tabular}
\end{center}
\caption{Black (gray) basis vectors are related to irreducible representations of the symmetric group that are (not) exterior powers of the fundamental representation. }
\label{tab:Vn}
\end{table}

\noindent{\it Acknowledgment}  I am grateful to Lisa Carbone, Hermann Nicolai, and Edward Witten for discussions.  This material is based upon work supported by the U.S. Department of Energy, Office of Science, Office of High Energy Physics under Award Number DE-SC0009988.

\bibliographystyle{jhep}
\bibliography{hyperbolic}

\end{document}